\documentclass[referee]{raa}            % referee version: for submission

\usepackage{graphicx,times}             %for PS/EPS graphics inclusion, new
\usepackage{natbib}
\usepackage{amssymb,amsmath}
\bibpunct{(}{)}{;}{a}{}{,}

\usepackage[a4paper=true,dvipdfm=true,pagebackref=true]{hyperref}
\hypersetup{colorlinks = true, linkcolor = green, anchorcolor = red, citecolor = blue, filecolor = red, pagecolor = red, urlcolor = red}
\def\hip{{\it Hipparcos}}
\def\akari{{\it AKARI}}
\def\fis{{\it AKARI/FIS}}
\def\iras{{\it IRAS}}
\def\iso{{\it ISO}}
\def\spiz{{\it Spitzer}}
\def\mass{{\it 2MASS}}
\def\wise{{\it WISE}}
\def\herschel{{\it Herschel}}
\def\alma{{\it ALMA}}

\begin{document}

   \title{A \wise\ View of \iras\ Debris Disks: Revising the Dust Properties}
%   \subtitle{I. Place Your Subtitle Here}

   \volnopage{Vol.0 (20xx) No.0, 000--000}      %%preserved for Editor. DOn't remove!
   \setcounter{page}{1}          %%starting page, preserved for Editor. DOn't remove!

   \author{Qiong Liu
      \inst{1}
   }

   \institute{Department of Physics, Guizhou University,
             Guiyang 550025, China; {\it qliu1@gzu.edu.cn}\\
\vs\no
   {\small Received~~20xx month day; accepted~~20xx~~month day}}

\abstract{
Debris disks around stars are considered as components of planetary systems. Constrain the dust properties of these disks can give crucial information to formation and evolution of planetary systems. As an all-sky survey, \textit{InfRared Astronomical Satellite} (\iras) gave great contribution to the debris disk searching which discovered the first debris disk host star (Vega). The \iras-detected debris disk sample published by Rhee \citep{rhe07} contains 146 stars with detailed information of dust properties. While the dust properties of 45 of them still can not be determined due to the limitations with the \iras\ database (have \iras\ detection at 60 $\mu$m only). Therefore, using more sensitivity data of \textit{Wide-field Infrared Survey Explorer} (\wise), we can better characterize the sample stars: For the stars with \iras\ detection at 60 $\mu$m only, we refit the excessive flux densities and obtain the dust temperatures and fractional luminosities; While for the remaining stars with multi-bands \iras\ detections, the dust properties are revised which show that the dust temperatures were over estimated in high temperatures band before.
Moreover, we identify 17 stars with excesses at the \wise\ 22 $\mu$m which have smaller distribution of distance from Earth and higher fractional luminosities than the other stars without mid-infrared excess emission. Among them, 15 stars can be found in previous works.
}
   \authorrunning{Q. Liu}

   \maketitle

\section{Introduction}           %% first-level sections will be auto-capitalized
\label{sect:intro}

Debris disks are almost dust-dominated and surround their stars with a wide age range \citep{hug18}. The dust is usually not primordial because its lifetime was much shorter than the stellar age \citep{wya18}. Debris disks are considered as planetary systems components or protoplanetary disks descendants, which provides a wealth of valuable information on evolution of circumstellar disks and planet formation outcome \citep{wya17}.

Since \iras\ first discovered Vega with debris disk in 1984 \citep{aum84}, attempts of searching debris disk candidates are never stop.
The most effective way of searching debris disk is to detect the infrared (IR) excess with IR radiation exceeds the stellar photospheric radiation %\add{\citep{bry06,su06}}.
The dust surrounds the sample stars and reaches a thermal equilibrium under stellar
radiation, then it will re-emits the light absorbed from the host star at IR to sub-millimeter \citep{kri10}.
%Therefor, searching debris disk candidate relies on infrared telescope.

Up to now, many hundreds of debris disks have been discovered  \citep{wya18}. They are
based on five satellites as follows:  \iras\ \citep{rhe07,man98},  \textit{Infrared Space Observatory} (\iso)
\citep{oud92,kes96,abr99,faj99,hab99,spa01,dec03}, \spiz\ \citep{chen05,rie05,kim05,su06,
bei06,moo06,bry06,sie07,reb08,moo11,wu12,chen14,mit15,bal18}, \herschel\ \citep{mat10,eir13,dod16,vic16,sib18} ,  and \akari\ \citep{fuj13,liu14,ish17}.
Among these surveys, \iras\ and \akari\ are all-sky surveys and the other three (\spiz, \iso\ and \herschel) are not. Though with much smaller area covering, the latter three missions have much better sensitivities and spatial resolutions so that they can detect more faint disks.

However, \iras\ and \akari\  still have their own advantages because of their all-sky area. And although both \iras\ and \akari\ are all-sky surveys, the debris disk candidates detected by them are not exactly the same.
The debris disk sample of Rhee \citep{rhe07} detected by \iras\ contains 146 stars with debris disk candidates via cross-correlating the \iras\ Point Source Catalog (PSC) and Faint Source Catalog (FSC) with the \hip\ main sequence star catalog.
And the debris disk sample of Liu \citep{liu14} detected by \akari\ contains 72 stars debris disk candidates via cross-correlating the AKARI/Far-Infrared Surveyor \citep{kaw07} All-Sky Survey Bright Source Catalogue (AKARIBSC, \citealt{yam10}) with  the \hip\ main sequence star catalog.
Among these two samples, 27 stars are in common although they have the similar sensitivity (\iras\ at 60 $\mu$m band and \fis\ at 90 $\mu$m). Most of the sources in Rhee's sample cannot detected by \fis\ because of its shallow limit.
So Rhee's sample still has its own value.

The sample of Rhee gives dust properties of debris disks including dust temperature, fractional luminosity and dust mass. It present a good sample for other works, such as follow up observations of debris disks (\herschel: \citealt{mar13,vic16}, \spiz: \citealt{chen14,mit15}, \alma: \citealt{boo19} and  Submillimeter: \citealt{bul13,hol17});
debris disks around A-type stars \citep{gre16,wel18,moo17};
Individual disk research \citep{fuj09,bor14,hun15,su15,kon16,gei19}.
Moreover, this sample focus on  debris disks evolution which gives great guidance to evolution works of other people \citep{wya07,moo11,vic14}. And this sample is also helpful to other statistical works: metallicity \citep{mal12}, binaries \citep{rod12}, Kuiper Belts \citep{nil10} and so on.

However, the dust properties of 45 of them still can not be determined  due to the limitations with the \iras\ database: these stars have only \iras\ 60 $\mu$m detection.
This limitation can be broken through with more sensitivity observation data of \textit{Wide-field Infrared Survey Explorer} (\wise).
Many studies have used \wise\ to search for IR excess stars \citep{cot16,wu13,wu16}.
\wise\ makes an all-sky survey at four IR bands with $W1$ at 3.4$\mu$m, $W2$ at 4.6$\mu$m, $W3$ at 12$\mu$m and $W4$  at 22 $\mu$m. The angular resolutions of corresponding bands are 6\farcs1, 6\farcs4, 6\farcs5 \& 12\farcs and the 5$\sigma$ point source sensitivities of corresponding bands are better than 0.08, 0.11, 1 and 6 mJy. For high SNR sources, the astrometry precision is better than 0\farcs15 \citep{wri10}.
The flux densities of $W1$ and $W2$ can be used to test the fitting quality of model spectra. And the flux densities of $W3$ and $W4$ can be used to fit the dust components and revise the properties.
%So the sensitivities at 12 \& 22 $\mu$m of \wise\ are far more sensitivity than the average 10 $\sigma$ sensitivities at 12 \& 25 $\mu$m of \iras\ which are 0.7 and 0.65 Jy only \citep{neu84}.

This paper has refitted the SEDs of Rhee's sample stars and refitted the excessive flux densities with \wise\ all-sky source catalog and \iras\ catalog to revise the dust properties and get Mid-IR excess information. The sample and photosphere emission are described in Section \ref{sect:sample} and results and analysis including the properties of debris disks and hosting stars as well as \wise\ 22$\mu$m excess are presented in Section \ref{sect:analysis}. I will discuss the revised dust properties and Mid-IR excess sample in Section \ref{sect:discussion} and draw the summary in Section \ref{sect:summary}.

\section{The Sample and Photosphere Emission}
\label{sect:sample}
\subsection{The Sample}

The sample used in this work is debris disks of 146 stars within 120 pc of Earth detected by \iras\ \citep{rhe07},
which cross-correlate the \iras\ Point Source Catalog (PSC) and Faint Source Catalog (FSC) with the \hip\ main sequence star catalog.
\iras\ makes an all-sky survey at four bands centered at 12, 25, 60, and 100 $\mu$m \citep{neu84}.
The \hip\ main sequence star catalog has more than 110,000 stars with the information of photometry and astrometry for the nearby stars \citep{bes00}.

All stars have debris disks around, which are identified by excesses at \iras\ 60 $\mu$m as described in the paper of Rhee. We will check these results in the Section 3.
Note that 3 stars were removed from the sample: 2 pre-main-sequence stars (HIP 53911, HIP 77542) and 1 star (HIP 19704) with a detection at 60 $\mu$m flux quality of 1 (which means the flux quality is not reliable). The sample has 143 stars left over.

\subsection{The Photosphere Emission}

It is essential to obtaining the 143 sample stars' photosphere flux densities in order to identify and measure the IR excess strength \citep{bry06}.
To construct the SEDs, we collected the optical data ($B$ and $V$ from the \hip\ satellite measurements)  to near-infrared (NIR) absolute photometric data ($JHK_s$ from Two Micron All Sky Survey (\mass) catalog) for all sample stars \citep{skr06} and converted these observed magnitudes into flux density (Janskys) by using
the zero magnitudes in \citealt{cox00}.
The photometry of the sample stars are listed in Table \ref{Tab:flux}.
Then we use Kurucz' models (ATLAS9,\citealt{cas04}) to fit the stellar SEDs as our previous work do \citep{liu14}.  The best model was selected out with the minimum $\chi^2$ which are presented in Table \ref{Tab:flux}.

\section{Results and Analysis}
\label{sect:analysis}

With the best-fit Kurucz model, we can estimate the flux densities of stellar photosphere in the corresponding \wise\ and \iras\ bands.
We check the \iras\ excesses of all 143 sample stars by using the criterion [F$_{IRAS}$ - F$_{phot}$] / $\sigma_{IRAS}$ $>$ 3.0,
where $F_{IRAS}$ is the \iras\ flux densities; $F_{phot}$ is the predicted photospheric
flux densities of corresponding bands; and $\sigma_{IRAS}$ is the uncertainties of the \iras\ flux densities in corresponding bands.

\subsection{\wise\ 22 $\mu$m Excess}
The observation data of \wise\ will be used to search Mid-IR excess objects.
All 143 stars are covered by \wise.
While our previous work showed that the model fluxes at \wise\ 12 and 22 $\mu$m have the systematical uncertainties $\sigma_{sys}$ with $\sigma_{w3}=0.06$ and $\sigma_{w4}=0.13$ \citep{liu14}.
Therefore, the systematical uncertainty should be considered to the uncertainties of \wise\ all bands $\sigma_{WISE}$ with $\sigma_{WISE}=$ $sqrt(\sigma_{obs}^2 + \sigma_{sys}^2)$, where $\sigma_{obs}$ means the observational uncertainties of \wise .
Note that 3 stars (HIP 70952, HIP 71284, HIP 74946) were removed from the sample
with [F$_{w4obs}$ - F$_{w4phot}$] $<$ $-1*\sigma_{w4}$ where F$_{w4obs}$ is the \wise\ 22 $\mu$m flux density and F$_{w4phot}$ is the predicted photospheric flux density at 22 $\mu$m band.
Therefore, there are 140 stars left in our sample for further discussion. These 140 stars's \wise\ magnitudes information can be seen from Table \ref{Tab:flux}.

The flux densities of $W1$ and $W2$ can be used to test the fitting quality of model spectra. And the flux densities of $W3$ and $W4$ can be used to fit the dust components and revise the dust properties of the stars in Rhee et al. (2007).
We can estimate the Mid-IR excesses at \wise\ 22 $\mu$m in the same way as 60 $\mu$m excess shown:

[F$_{w4obs}$ - F$_{w4phot}$] / $\sigma_{w4}$ $>$ 3.0,

We identify 31 stars with excesses at the 22 $\mu$m by applying this criterion.
While this criterion is affected by the \iras\ 100 $\mu$m background, whose level should be lower than 5 MJy/sr as \citealt{ken12} and \citealt{wu13} shown. We check the 31 $\mu$m excess stars and find indeed there are 14 stars may be affected. So there are only 17 stars left after this cut.
Mid-IR excess stars are thought to have co-existence of hot and cold dust components just like our solar system. These 17 stars are putted into Mid-IR excess sample which are listed in Table \ref{Tab:WISE}.
At the following sections, I will discuss this two sub-samples: Mid-IR excess sample and Non Mid-IR excess sample.

\subsection{Properties of Debris Disk Host stars}

\begin{figure}
   \centering
   \includegraphics[width=\textwidth, angle=0]{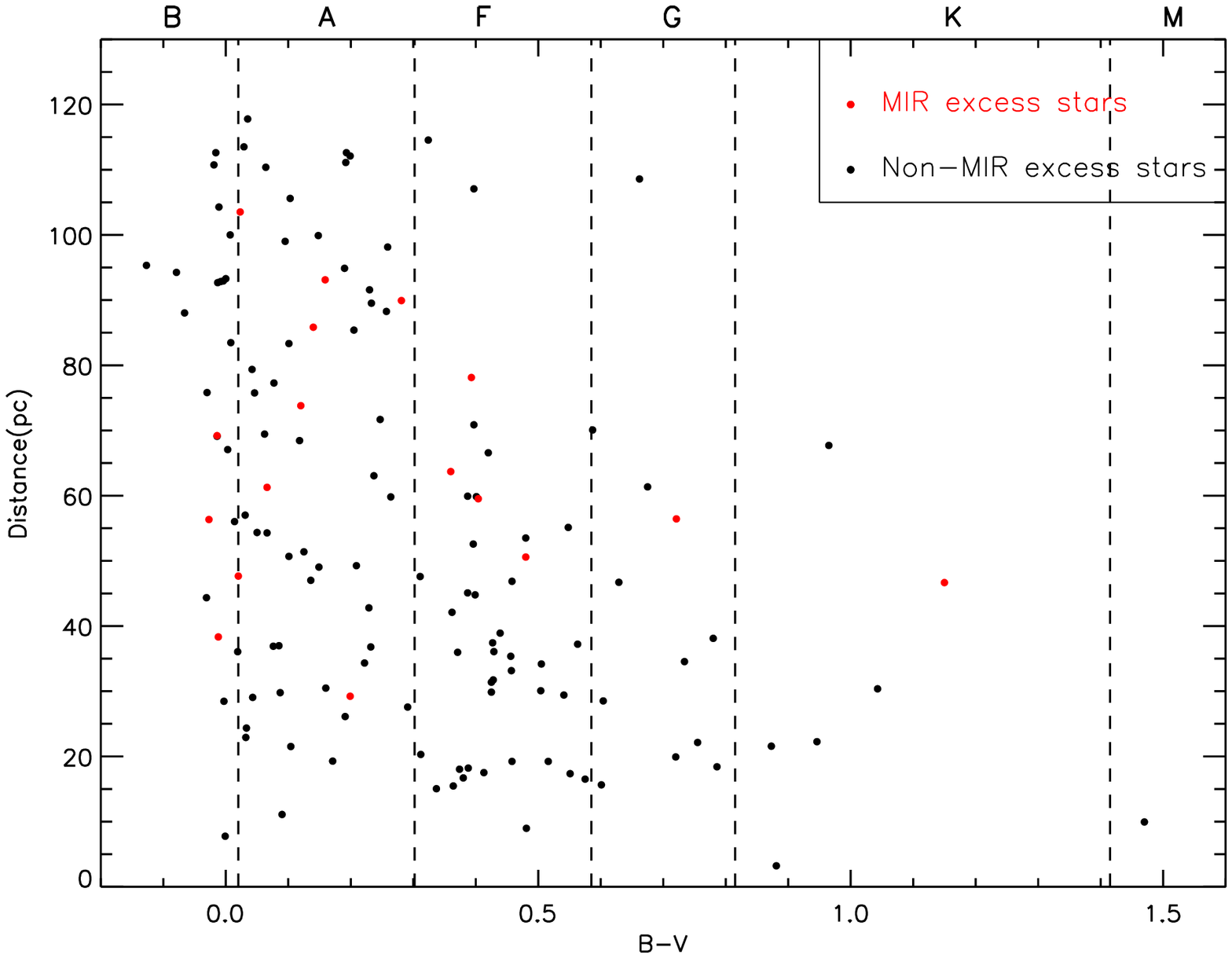}
   \caption{Distribution of Mid-IR excess sample and Non Mid-IR excess sample stars in distance from Earth as a function of B-V.}
   \label{Fig1}
   \end{figure}

In this subsection, I will study the stellar properties of debris disk hosts including color, distance from the Earth and location on the H-R diagram.
Debris disk host stars are more inclined to early type stars as Rhee et al. (2007) pointed out which can be seen from the figure of sample stars function distribution of the distance from Earth and B-V. We re-draw the figure and mark the Mid-IR excess objects in red dots as Figure \ref{Fig1} shown.
From the figure, we can see Mid-IR excess sample has smaller distribution of distance from Earth (29.2 pc to 103.5 pc) than Non Mid-IR excess sample stars (3.2 pc to 117.8 pc). Note that there are only 17 Mid-IR excess stars, the sample is so small that may lead to a false distribution trend.

\subsection{Dust Properties}

In this subsection, I will study the dust properties including the dust temperatures and fractional luminosities of debris disks.

\subsubsection{Dust Temperatures}
Debris disks usually consist of a single narrow ring which reach thermal equilibrium in the field of stellar radiation as previous studies suggested \citep{bac93}. Rhee et al. (2007) used the blackbody model with single-temperature to fit the dust component and obtained dust temperature $T$ for each star.

Since \iras\ sample of Rhee is based on solely \iras, the dust properties of 45 objects in their sample still can not be determined due to the limitations with the \iras\ database (have \iras\ detection at 60 $\mu$m only). They set the dust temperature to 85 K so the peak will fall at 60 $\mu$m. According to the number of the \iras\ detected bands, I divide the 140 stars sample to 2 groups:

(1) Group I contains 45 stars with only \iras\ 60 $\mu$m detection;

(2) Group II contains the remaining 95 stars with multi-bands \iras\ detections.

\begin{figure}
   \centering
   \includegraphics[width=\textwidth, angle=0]{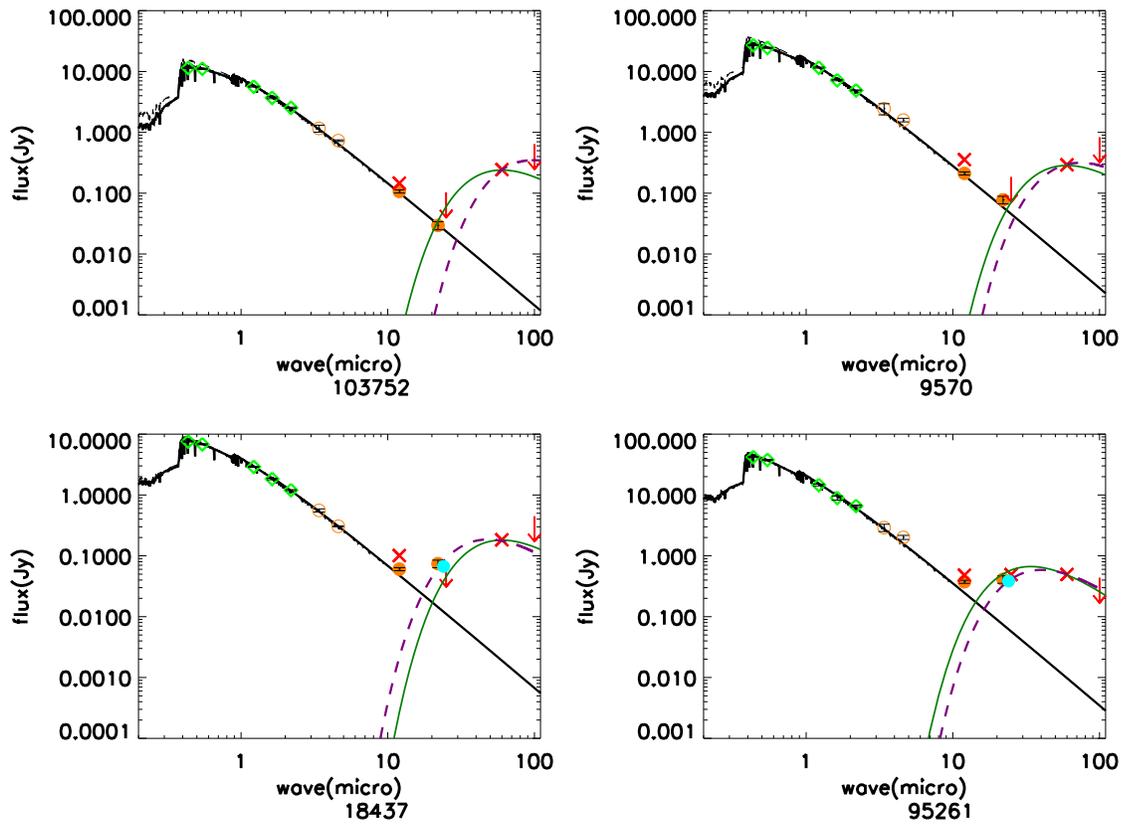}
   \caption{ SEDs for 4 IR excess stars. The photospheric models and the disk models are shown as solid black and dotted purple lines, respectively. The green solid lines are the disk models of Rhee. The different symbols represent the different data sets: green diamonds, BVJHK; red, IRAS; cyan filled dots, Spitzer; orange filled dots, WISE without saturations; and orange hollow circles, WISE with saturations.}
   \label{Fig2}
   \end{figure}

The excessive flux densities are fitted in IR bands including 60, 100 $\mu$m of \iras\ and 12, 22 $\mu$m of \wise\ with a blackbody model of single temperature.
The SED fittings of 4 stars as an example are presented in Figure \ref{Fig2}.
From SED fittings, the dust temperatures can be derived
with group I stars listed in Table \ref{Tab:groupi} and Group II stars listed in Table \ref{Tab:groupii}.
From the figure, one can easily see that our fitting results are better than Rhee's because of the addition of new \wise\ data.

\subsubsection{Fractional Luminosities}

From the SED fitting, we can also estimate the fractional luminosity $f$ which is used to characterize the disk's effective optical depth.
The fractional luminosity $f$ is calculate by divided the IR luminosity of debris disk to the stellar luminosity (\citealt{wya08}),
\begin{equation}
f = L_{\rm ir}/ L_\star
\end{equation}
where $L_\star$ is the stellar luminosity estimated by the best model of SED fitting.
The IR luminosity $L_{\rm ir}$ is calculated from the blackbody model of fitted IR.
The calculated fractional luminosities for all sample stars are listed in Table \ref{Tab:groupi} and Table \ref{Tab:groupii} as well.

\section{Discussion}
\label{sect:discussion}

In this section, I will first give my discussion on revising the dust properties including dust temperatures and fractional luminosities. And then, I will discuss the Mid-IR excess sample.

\subsection{Revising the Dust Temperatures}

The revised dust temperature can be seen in Table \ref{Tab:groupi} and Table \ref{Tab:groupii}.
For Group I stars, we can get the dust temperatures which were set to 85 K in \citealt{rhe07}.
And for Group II stars,  even they have multi-bands \iras\ detections,
the dust temperatures of ours are well determined than that in \citealt{rhe07} because of the much better sensitivity in the Mid-IR of \wise\ in comparison with \iras.

\begin{figure}
   \centering
   \includegraphics[width=\textwidth, angle=0]{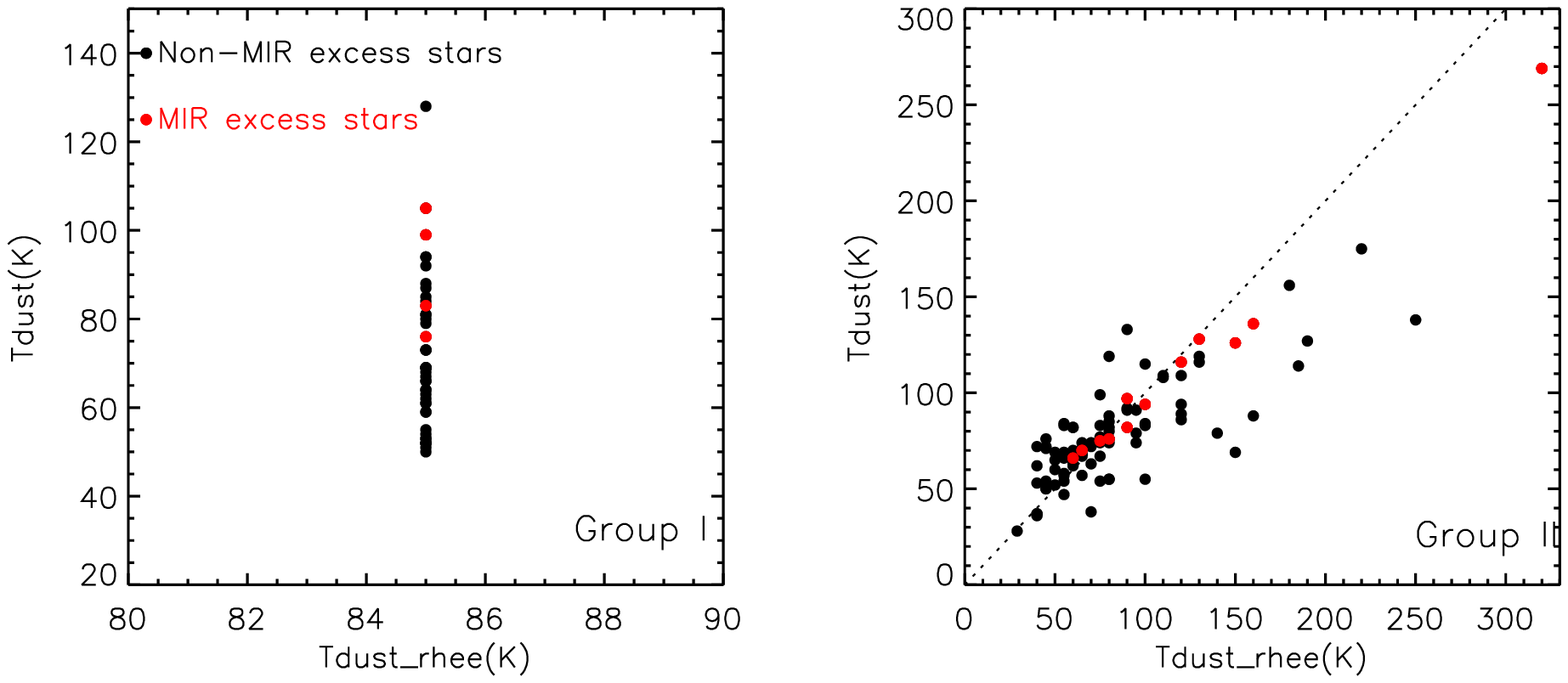}
   \caption{The differences of dust temperatures between Rhee et al. (2007) and ours. Group I and Group II stars are shown in the left and right panels, respectively. The different symbols represent the different sub-sample stars: red filled dots are MIR excess stars and black filled dots are the Non-MIR excess stars.}
   \label{Fig3}
   \end{figure}

The difference of dust temperatures between Rhee et al. (2007) and ours is shown in Figure \ref{Fig3}.
From Figure \ref{Fig3}, we can see the dust temperatures of Group II stars are over estimated in high temperatures band.
While in lower temperature band ($<$100 K), dust temperatures have a part under estimated and a part over estimated, which have no obvious favoritism.

\subsection{Revising the Fractional Luminosities}

\begin{figure}
   \centering
   \includegraphics[width=\textwidth, angle=0]{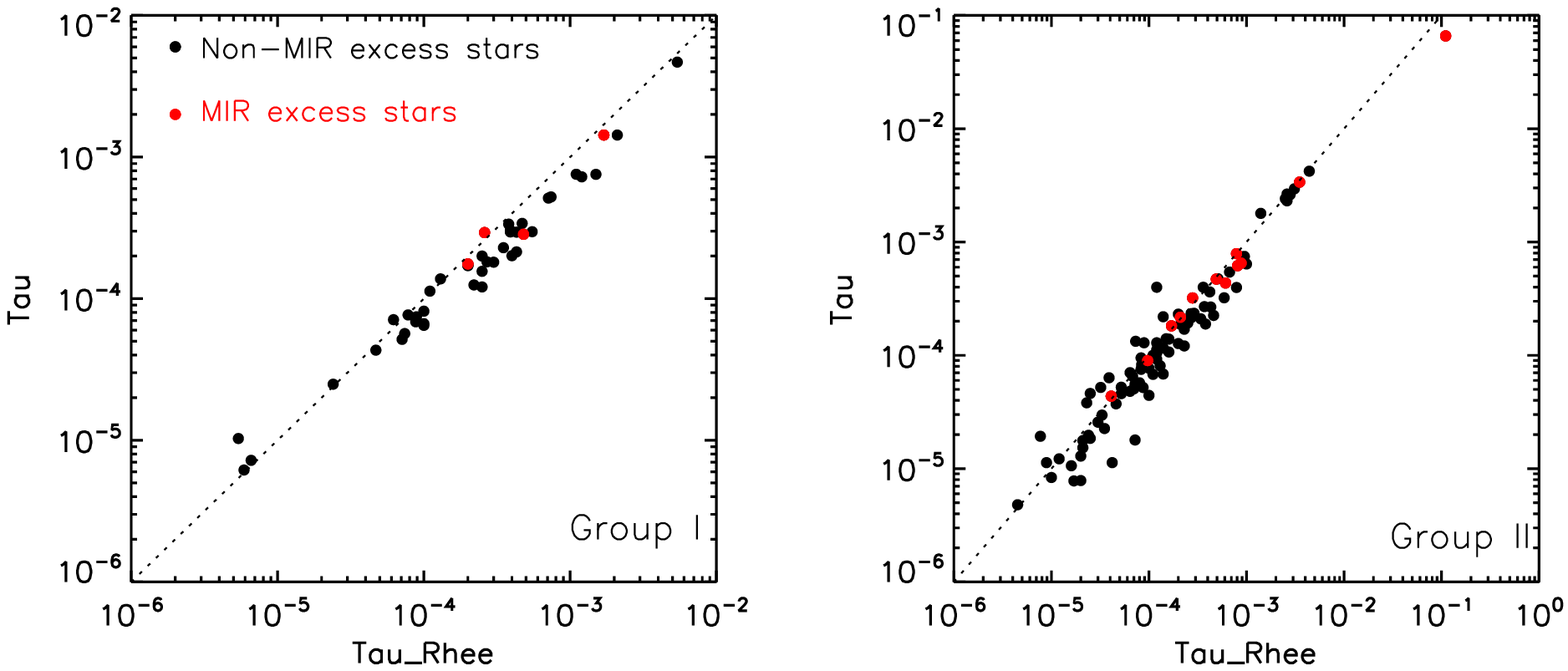}
   \caption{The differences of fractional luminosities between Rhee et al. (2007) and ours. Group I and Group II stars are shown in the left and right panels, respectively. The different symbols represent the different sub-sample stars: red filled dots are MIR excess stars and black filled dots are the Non-MIR excess stars.}
   \label{Fig4}
   \end{figure}

The differences of fractional luminosities $f$ between Rhee et al. (2007) and ours are shown in Figure \ref{Fig4}.
It is obvious that the fractional luminosities of Group I stars are high estimated by Rhee due to the high estimated dust temperatures which can be seen from the left panel of Fig. \ref{Fig4}.
While for Group II stars, the fractional luminosities show no obvious favoritism with a part under estimated and a part over estimated which can be seen from the right panel of Fig. \ref{Fig4}. However, for the Mid-IR excess sample stars in this group, fractional luminosities of mine are almost the same as Rhee et al. (2007). It is most likely because they have plenty of IR excess data to do blackbody fittings.

Whether in Group I or Group II, Mid-IR sample stars have high fractional luminosities than Non Mid-IR sample stars which imply that the stars with higher fractional luminosities may have higher probability of having warm dust component.

\subsection{Mid-IR Excess Sample (\wise\ 22 $\mu$m)}

From Section \ref{sect:analysis}, Mid-IR excess sample contains 17-stars with \wise\ 22 $\mu$m excesses.
Such stars with Mid-IR excess emission may have terrestrial planets \citep{pad16}.

Up to now, a few hundreds of warm disks with Mid-IR excess have been discovered with \spiz, \akari\ and
\wise\ \citep{mey08,fuj10a,fuj10b,olo12,rib12,fuj13,wu12,wu13,liu14,ish17,bal18}, and such disks' incidence decreases very rapidly with increasing stellar ages \citep{urb12}.

What is the cause of the warm component? For most warm debris disks, the Mid-IR excesses could be explained by giant impact stages (about 100 Myr) \citep{gen15}. They found that, after a giant impact, the IR excess is sometimes almost 10 times higher than the stellar IR flux.

In Mid-IR excess sample, 15 stars can be found in previous work as Table \ref{Tab:WISE} column 3 shown.
Note \add{that} maybe the remaining 2 stars can be seen in the other work elsewhere.

\section{Summary}
\label{sect:summary}

This paper has refitted the SEDs of the sample stars with the Kurucz' models and refitted the excessive flux densities with \wise\ and \iras\ all sky catalogs.
we obtain the dust temperatures of Group I stars with only 60 $\mu$m data which can not determined in previous study of Rhee.
And for Group II stars,  even they have \iras\ multi-bands detections,
the dust temperatures of ours are well determined than that in \citealt{rhe07} because of the much better sensitivity in the Mid-IR of \wise\ in comparison with \iras.
From the revised dust properties, we can see that the dust temperatures of Group II stars were over estimated in high temperatures band before and the fractional luminosities of Group I stars were high estimated.

Moreover, we identify 17 stars with \wise\ 22 $\mu$m excess and discuss the difference from Non Mid-IR sample.
Disks around the Mid-IR sample stars appear to be more bright and with higher dust temperatures.
We hope these revision of dust properties can give some guidance to the follow-up works.
%After adding other available data of other telescopes such as \spiz, \herschel,
%the dust properties of some stars in these sample will be constrained better which can be done in our future works.

\begin{acknowledgements}

I am grateful to the anonymous referee for his/ her comments that improved the paper.
This work was supported by the National Natural Science Foundation of China (Grant No.U1631109).
This work is based on the sample of Rhee and makes use of data products from many telescopes:
\wise\ (a joint project of the University of California, Los Angeles, and the Jet Propulsion Laboratory/California Institute of Technology), \hip\ ( the primary result of the Hipparcos space astrometry mission, undertaken by the European Space Agency) and \mass\ (a joint project of the University of Massachusetts and the Infrared Processing and Analysis Center /California Institute of Technology).
This research makes use of ATLAS9 model and the SIMBAD database, operated at the
CDS, Strasbourg, France. And this work makes use of the NASA/IPAC Infrared Science Archive,
which is operated by the Jet Propulsion Laboratory, California Institute of Technology,
under contract with the National Aeronautics and Space Administration.

\end{acknowledgements}

%-------------------------------------------------------------
\clearpage

\clearpage

\clearpage

\clearpage

%-------------------------------------------------------------
\clearpage
\begin{table}
\begin{center}
%\tiny
\caption[]{Photometry and Flux Density for the 140 Sample Stars. }\label{Tab:flux}
 \begin{tabular}{rcccccccccccccccccccl}
  \hline\noalign{\smallskip}
       HIP   & B & V & J & H & K & w1m & w2m &  w3m &  w4m & f12 & f25& f60& f100 & $\chi^2$ \\
          & mag & mag & mag & mag & mag & mag & mag & mag & mag & Jy & Jy & Jy & Jy &  \\
        \hline\noalign{\smallskip}
        746  &  2.66  &  2.28 &   1.71 &   1.58  &  1.45 & -0.876 & -0.178 &  1.462 &  1.335 &  11.700  &   2.871  &   1.001  &  12.660 & 5.85e-01 \\
     1185  &  7.08  &  6.82 &   6.38 &   6.31  &  6.25 &  6.191 &  6.142 &  6.284 &  6.228 &   0.175  &   0.167  &   0.211  &   0.567 & 4.46e-03 \\
     4267  &  5.78  &  5.80 &   5.74 &   5.80  &  5.80 &  5.801 &  5.734 &  5.825 &  5.454 &   0.193  &   0.149  &   0.174  &   1.309 & 1.00e-03 \\
     5626  &  5.61  &  5.60 &   5.46 &   5.50  &  5.49 &  5.535 &  5.388 &  5.493 &  4.656 &   0.260  &   0.111  &   0.403  &   2.081 & 3.22e-02 \\
     6686  &  2.82  &  2.66 &   2.34 &   2.37  &  2.25 &  0.834 &  0.793 &  2.378 &  2.286 &   4.961  &   1.142  &   0.346  &   6.092 & 2.11e+00 \\
     6878  &  7.17  &  6.66 &   5.69 &   5.49  &  5.42 &  5.443 &  5.256 &  5.441 &  5.321 &   0.310  &   0.112  &   0.285  &   1.045 & 2.60e-01 \\
     7345  &  5.69  &  5.62 &   5.49 &   5.53  &  5.46 &  5.471 &  5.302 &  5.340 &  3.742 &   0.335  &   0.384  &   2.018  &   1.883 & 4.48e+00 \\
     7805  &  8.04  &  7.62 &   6.84 &   6.69  &  6.63 &  6.619 &  6.592 &  6.598 &  6.040 &   0.119  &   0.093  &   0.140  &   0.646 & 1.55e-01 \\
     7978  &  6.07  &  5.52 &   4.79 &   4.40  &  4.34 &  4.171 &  3.913 &  4.220 &  3.954 &   0.814  &   0.282  &   0.815  &   1.075 & 1.04e+01 \\
     8122  &  6.98  &  6.73 &   6.26 &   6.19  &  6.17 &  6.154 &  6.100 &  6.183 &  5.640 &   0.141  &   0.107  &   0.348  &   0.607 & 1.49e-02 \\
     8241  &  5.07  &  5.04 &   4.99 &   5.03  &  4.96 &  4.932 &  4.717 &  4.984 &  4.594 &   0.405  &   0.163  &   0.299  &   0.616 & 4.44e+00 \\
     9570  &  5.53  &  5.50 &   5.35 &   5.38  &  5.33 &  5.246 &  5.078 &  5.345 &  5.069 &   0.356  &   0.184  &   0.293  &   0.829 & 2.03e-01 \\
    10054  &  6.17  &  6.05 &   5.75 &   5.76  &  5.69 &  5.741 &  5.607 &  5.695 &  5.523 &   0.206  &   0.050  &   0.153  &   1.161 & 1.52e-01 \\
    10670  &  4.05  &  4.03 &   3.80 &   3.86  &  3.96 &  3.952 &  3.643 &  3.989 &  3.510 &   1.066  &   0.367  &   0.791  &   0.831 & 2.86e+00 \\
    11360  &  7.19  &  6.79 &   6.03 &   5.86  &  5.82 &  5.814 &  5.646 &  5.778 &  5.269 &   2.130  &   0.868  &   0.441  &   0.663 & 6.70e-01 \\
    11486  &  5.60  &  5.29 &   5.08 &   4.82  &  4.59 &  4.574 &  4.380 &  4.618 &  4.381 &   0.549  &   0.192  &   0.282  &   0.655 & 6.85e+00 \\
    11847  &  7.83  &  7.47 &   6.70 &   6.61  &  6.55 &  6.544 &  6.520 &  6.503 &  4.236 &   0.125  &   0.177  &   0.868  &   0.853 & 8.35e-01 \\
    12361  &  7.17  &  6.78 &   6.11 &   5.97  &  5.92 &  5.955 &  5.799 &  5.904 &  5.406 &   0.157  &   0.086  &   0.304  &   0.768 & 6.26e+00 \\
    12964  &  6.87  &  6.48 &   5.76 &   5.63  &  5.57 &  5.554 &  5.357 &  5.585 &  5.457 &   0.253  &   0.140  &   0.226  &   0.630 & 1.00e-01 \\
    13005  &  9.03  &  8.06 &   6.42 &   5.99  &  5.88 &  5.887 &  5.842 &  5.913 &  5.859 &   0.138  &   0.245  &   0.288  &   1.013 & 1.24e+00 \\
    13141  &  5.35  &  5.25 &   5.14 &   5.16  &  4.97 &  4.963 &  4.748 &  5.012 &  4.781 &   0.414  &   0.097  &   0.164  &   0.490 & 1.66e+00 \\
    14576  &  2.09  &  2.09 &   1.96 &   1.95  &  1.89 & -0.011 &  0.244 &  1.903 &  1.782 &   7.348  &   1.866  &   0.455  &   1.685 & 5.20e-01 \\
    15197  &  5.03  &  4.80 &   4.42 &   4.25  &  4.22 &  4.149 &  3.869 &  4.135 &  4.102 &   0.848  &   0.235  &   0.214  &   1.566 & 4.04e+00 \\
    16449  &  6.50  &  6.38 &   6.16 &   6.12  &  6.10 &  6.097 &  6.045 &  6.115 &  5.372 &   0.173  &   0.101  &   0.595  &   0.636 & 1.04e-01 \\
    16537  &  4.60  &  3.72 &   2.23 &   1.88  &  1.78 &  2.970 &  2.285 &  1.770 &  1.288 &   9.672  &   2.667  &   1.631  &   2.008 & 4.95e+00 \\
    18437  &  6.91  &  6.89 &   6.85 &   6.87  &  6.86 &  6.851 &  6.868 &  6.705 &  5.118 &   0.101  &   0.080  &   0.183  &   0.446 & 1.98e+00 \\
    18859  &  5.90  &  5.38 &   4.71 &   4.34  &  4.18 &  4.146 &  3.926 &  4.289 &  3.882 &   0.982  &   0.247  &   0.287  &   2.242 & 7.76e-02 \\
    18975  &  5.82  &  5.45 &   4.81 &   4.58  &  4.51 &  4.529 &  4.269 &  4.530 &  4.442 &   0.639  &   0.262  &   0.209  &   1.923 & 6.27e-01 \\
    19704  &  7.31  &  6.99 &   6.34 &   6.29  &  6.18 &  6.093 &  5.945 &  6.160 &  6.060 &   0.131  &   0.050  &   0.239  &   0.938 & 1.78e+00 \\
    19893  &  4.57  &  4.26 &   3.68 &   3.47  &  3.51 &  3.487 &  3.188 &  3.497 &  3.399 &   1.661  &   0.441  &   0.321  &   1.000 & 1.07e-01 \\
    20635  &  4.35  &  4.21 &   4.09 &   4.06  &  4.08 &  3.929 &  3.563 &   null &  3.830 &   1.115  &   0.310  &   0.353  &   4.666 & 5.30e-04 \\
    21604  &  5.83  &  5.85 &   5.66 &   5.59  &  5.56 &  5.758 &  5.515 &  5.603 &  5.391 &   0.235  &   0.173  &   0.637  &   7.072 & 1.97e-02 \\
    22226  &  8.24  &  7.85 &   7.10 &   6.95  &  6.89 &  6.872 &  6.867 &  6.867 &  6.000 &   0.092  &   0.066  &   0.274  &   0.478 & 1.23e-01 \\
    22439  &  6.73  &  6.27 &   5.38 &   5.20  &  5.08 &  5.102 &  4.927 &  5.146 &  4.997 &   0.387  &   0.091  &   0.191  &   1.011 & 1.14e-01 \\
    22845  &  4.72  &  4.64 &   4.85 &   4.52  &  4.42 &  4.414 &  4.166 &  4.432 &  4.058 &   0.696  &   0.270  &   0.411  &   5.217 & 5.62e+00 \\
    23451  &  8.33  &  8.13 &   7.69 &   7.62  &  7.59 &  7.595 &  7.629 &  6.924 &  3.952 &   0.114  &   0.211  &   1.122  &   1.592 & 1.17e-29 \\
    24528  &  6.90  &  6.76 &   6.50 &   6.43  &  6.43 &  6.408 &  6.371 &  6.357 &  5.589 &   0.144  &   0.060  &   0.108  &   0.510 & 1.18e+00 \\
    25197  &  5.23  &  5.24 &   5.66 &   5.16  &  5.16 &  5.154 &  5.006 &  5.154 &  4.925 &   0.352  &   0.172  &   0.172  &   1.549 & 7.30e+00 \\
    25790  &  6.03  &  5.93 &   5.61 &   5.60  &  5.55 &  5.558 &  5.439 &  5.600 &  5.543 &   0.243  &   0.170  &   0.396  &   3.694 & 3.14e-01 \\
    26453  &  7.66  &  7.26 &   6.47 &   6.29  &  6.28 &  6.263 &  6.186 &  6.215 &  5.317 &   0.124  &   0.117  &   0.128  &   0.406 & 3.09e-01 \\
    26966  &  5.72  &  5.73 &   5.79 &   5.84  &  5.78 &  5.806 &  5.728 &  5.631 &  4.559 &   0.195  &   0.110  &   0.313  &   0.511 & 4.90e+00 \\
    27072  &  4.07  &  3.59 &   2.80 &   2.61  &  2.51 &  3.085 &  2.392 &  2.261 &  2.394 &   4.396  &   0.981  &   0.228  &   0.483 & 2.42e-30 \\
    27288  &  3.65  &  3.55 &   3.39 &   3.31  &  3.29 &  3.347 &  2.991 &  3.221 &  2.388 &   2.115  &   1.157  &   0.365  &   1.001 & 1.94e-01 \\
    27321  &  4.02  &  3.85 &   3.67 &   3.54  &  3.53 &  3.484 &  3.182 &  2.597 &  0.014 &   3.459  &   9.047  &  19.900  &  11.280 & 1.61e+00 \\
       \noalign{\smallskip}\hline
\end{tabular}
\end{center}
\end{table}

\clearpage
\begin{table}
\begin{center}
%\tiny
%\subcaption{\autoref{Tab:flux} - continued}
\subcaption{Table 1 - continued}
%\contcaption{Basic Information and sources}
 \begin{tabular}{rcccccccccccccccccccl}
  \hline\noalign{\smallskip}
       HIP   & B & V & J & H & K & w1m & w2m &  w3m &  w4m & f12 & f25& f60& f100 & $\chi^2$ \\
               & mag & mag & mag & mag & mag & mag & mag & mag & mag & Jy & Jy & Jy & Jy &  \\
        \hline\noalign{\smallskip}
    27980  &  8.28  &  7.65 &   6.49 &   6.23  &  6.15 &  6.169 &  6.069 &  6.170 &  6.084 &   0.169  &   0.150  &   0.940  &  14.910 & 2.51e-02 \\
    28103  &  4.05  &  3.71 &   3.06 &   2.98  &  2.99 &  2.021 &  1.749 &  2.905 &  2.792 &   2.878  &   0.726  &   0.212  &   1.648 & 3.72e-01 \\
    28230  &  7.83  &  7.55 &   7.01 &   6.93  &  6.88 &  6.851 &  6.853 &  6.834 &  5.708 &   0.087  &   0.061  &   0.193  &   0.435 & 3.54e-01 \\
    32480  &  5.82  &  5.24 &   4.41 &   4.07  &  4.13 &  3.872 &  3.439 &  3.954 &  3.825 &   1.463  &   0.412  &   0.381  &   1.020 & 5.04e+00 \\
    32775  &  6.57  &  6.11 &   5.30 &   5.13  &  5.02 &  5.003 &  4.794 &  5.053 &  4.990 &   0.391  &   0.084  &   0.183  &   1.643 & 6.71e-02 \\
    33690  &  7.60  &  6.81 &   5.46 &   5.10  &  4.99 &  4.959 &  4.698 &  4.980 &  4.672 &   0.400  &   0.267  &   0.180  &   1.137 & 4.66e-02 \\
    34276  &  6.51  &  6.52 &   6.49 &   6.48  &  6.48 &  6.486 &  6.411 &  6.489 &  5.853 &   0.124  &   0.097  &   0.203  &   0.940 & 1.73e-01 \\
    34819  &  6.25  &  5.85 &   5.20 &   4.98  &  4.83 &  4.904 &  4.661 &  4.877 &  4.771 &   0.551  &   0.109  &   0.151  &   0.699 & 7.91e-01 \\
    35550  &  3.87  &  3.50 &   2.82 &   2.60  &  2.56 &  1.491 &  1.004 &  2.622 &  2.526 &   3.903  &   0.834  &   0.256  &   0.622 & 7.28e-02 \\
    36906  &  8.34  &  7.68 &   6.49 &   6.23  &  6.16 &  6.135 &  6.036 &  6.145 &  6.053 &   0.176  &   0.135  &   0.145  &   0.567 & 1.44e-05 \\
    36948  &  8.96  &  8.23 &   6.91 &   6.58  &  6.46 &  6.448 &  6.369 &  6.416 &  5.644 &   0.357  &   0.250  &   0.613  &   9.612 & 5.86e-01 \\
    39757  &  3.29  &  2.83 &   2.17 &   2.08  &  2.02 & -0.227 & -0.061 &  1.961 &  1.856 &   7.199  &   1.757  &   0.397  &   1.145 & 2.84e-01 \\
    40938  &  7.64  &  7.24 &   6.34 &   6.15  &  6.09 &  6.089 &  5.932 &  6.097 &  5.986 &   0.168  &   0.149  &   0.165  &   1.297 & 3.08e-03 \\
    41152  &  5.64  &  5.52 &   5.25 &   5.29  &  5.25 &  5.255 &  5.092 &  5.270 &  4.838 &   0.372  &   0.140  &   0.177  &   0.494 & 5.09e-02 \\
    41307  &  3.90  &  3.91 &   4.12 &   4.09  &  4.08 &  3.927 &  3.556 &  3.925 &  3.429 &   1.087  &   0.394  &   0.280  &   1.874 & 2.86e+00 \\
    42028  &  5.81  &  5.80 &   5.77 &   5.82  &  5.73 &  5.795 &  5.707 &  5.819 &  5.697 &   0.202  &   0.129  &   0.144  &   0.593 & 4.20e-04 \\
    42430  &  5.77  &  5.05 &   4.05 &   3.59  &  3.59 &  3.446 &  3.183 &  3.480 &  3.405 &   1.784  &   0.380  &   0.151  &   0.549 & 3.10e+00 \\
    43970  &  5.37  &  5.22 &   4.91 &   4.97  &  4.87 &  4.894 &  4.527 &  4.927 &  4.593 &   0.470  &   0.223  &   0.319  &   0.796 & 1.97e-01 \\
    44001  &  5.89  &  5.68 &   5.27 &   5.21  &  5.16 &  5.158 &  4.978 &  5.195 &  4.886 &   0.324  &   0.235  &   0.391  &   1.386 & 9.43e-02 \\
    45758  &  6.88  &  6.62 &   6.15 &   6.04  &  5.96 &  6.032 &  5.886 &  6.015 &  5.917 &   0.216  &   0.118  &   0.192  &   0.586 & 1.44e-04 \\
    48164  &  7.43  &  7.20 &   6.75 &   6.68  &  6.63 &  6.637 &  6.611 &  6.666 &  6.177 &   0.119  &   0.136  &   0.243  &   0.628 & 1.60e-01 \\
    48541  &  7.75  &  7.59 &   7.21 &   7.20  &  7.19 &  7.095 &  7.164 &  7.103 &  6.071 &   0.113  &   0.128  &   0.193  &   0.487 & 7.59e-01 \\
    51438  &  4.76  &  4.72 &   4.75 &   4.71  &  4.57 &  4.555 &  4.348 &  4.609 &  4.531 &   0.606  &   0.155  &   0.145  &   1.092 & 2.11e+00 \\
    51658  &  4.94  &  4.72 &   4.12 &   4.06  &  4.20 &  4.121 &  3.875 &  4.210 &  4.169 &   0.919  &   0.185  &   0.397  &   0.955 & 3.04e+00 \\
    52462  &  8.59  &  7.72 &   6.18 &   5.77  &  5.66 &  5.613 &  5.509 &  5.629 &  5.556 &   0.219  &   0.083  &   0.248  &   0.984 & 8.08e-02 \\
    53524  &  7.59  &  7.36 &   6.91 &   6.87  &  6.79 &  6.717 &  6.745 &  6.692 &  5.475 &   0.281  &   0.250  &   0.601  &   1.830 & 2.01e+00 \\
    53910  &  2.37  &  2.34 &   2.27 &   2.36  &  2.29 &  1.156 &  0.845 &  2.461 &  2.138 &   4.795  &   1.387  &   0.627  &   1.000 & 3.98e+00 \\
    53911  & 11.64  & 10.92 &   8.22 &   7.56  &  7.30 &  7.009 &  6.878 &  4.540 &  1.516 &   0.613  &   2.435  &   4.105  &   4.900 & 1.78e+00 \\
    55505  & 10.04  &  8.89 &   6.40 &   5.76  &  5.59 &  5.496 &  5.344 &  3.110 &  0.202 &   1.981  &   9.443  &   7.895  &   4.565 & 5.34e-02 \\
    56253  &  6.38  &  6.12 &   5.66 &   5.60  &  5.56 &  5.589 &  5.435 &  5.604 &  5.435 &   0.243  &   0.114  &   0.141  &   0.863 & 3.59e-01 \\
    56675  &  6.00  &  5.64 &   4.97 &   4.92  &  4.70 &  4.718 &  4.483 &  4.698 &  4.606 &   0.568  &   0.250  &   0.275  &   7.545 & 8.22e-01 \\
    57632  &  2.23  &  2.14 &   1.85 &   1.93  &  1.88 &  0.461 &  0.129 &  2.056 &  1.698 &   7.067  &   2.323  &   1.027  &   1.120 & 7.91e+00 \\
    60074  &  7.64  &  7.04 &   5.87 &   5.61  &  5.54 &  5.522 &  5.357 &  5.540 &  5.182 &   0.259  &   0.135  &   0.705  &   0.909 & 1.73e-01 \\
    61174  &  4.69  &  4.30 &   3.61 &   3.37  &  3.37 &  2.685 &  2.912 &  3.308 &  2.752 &   2.242  &   0.770  &   0.308  &   0.803 & 1.56e-01 \\
    61498  &  5.78  &  5.78 &   5.78 &   5.79  &  5.77 &  5.368 &  5.399 &  5.024 &  1.222 &   0.433  &   3.379  &   7.359  &   3.814 & 1.69e+00 \\
    61782  &  8.14  &  7.99 &   7.64 &   7.59  &  7.58 &  7.523 &  7.564 &  7.053 &  3.913 &   0.094  &   0.266  &   0.368  &   1.155 & 1.53e-29 \\
    61960  &  4.96  &  4.88 &   4.99 &   4.76  &  4.68 &  4.649 &  4.298 &  4.700 &  4.259 &   0.547  &   0.235  &   0.270  &   0.579 & 2.64e+00 \\
    63584  &  6.44  &  6.01 &   5.19 &   5.05  &  5.00 &  4.983 &  4.757 &  5.033 &  4.914 &   0.362  &   0.150  &   0.130  &   0.424 & 6.06e-01 \\
    64375  &  6.68  &  6.49 &   6.12 &   6.02  &  6.01 &  5.980 &  5.906 &  6.014 &  5.631 &   0.196  &   0.182  &   0.330  &   0.796 & 6.75e-02 \\
    64921  &  7.28  &  7.07 &   6.65 &   6.59  &  6.57 &  6.634 &  6.303 &  6.435 &  6.529 &   0.116  &   0.062  &   0.192  &   1.779 & 3.37e+00 \\
    68101  &  6.78  &  6.00 &   4.57 &   4.02  &  3.99 &  4.008 &  3.792 &  4.020 &  3.965 &   1.047  &   0.194  &   0.382  &  10.230 & 1.81e-01 \\
    68593  &  7.72  &  7.16 &   6.16 &   5.94  &  5.88 &  5.882 &  5.794 &  5.883 &  5.701 &   0.187  &   0.079  &   0.150  &   0.305 & 1.85e-04 \\
    69682  &  9.56  &  8.89 &   7.65 &   7.38  &  7.27 &  7.166 &  7.250 &  7.230 &  6.678 &   0.133  &   0.128  &   0.207  &   0.404 & 5.72e-01 \\
    69732  &  4.27  &  4.18 &   3.98 &   4.03  &  3.91 &  3.922 &  3.758 &  4.118 &  3.611 &   1.151  &   0.356  &   0.472  &   1.000 & 3.74e+00 \\
       \noalign{\smallskip}\hline
\end{tabular}
\end{center}
\end{table}

\clearpage
\begin{table}
\begin{center}
%\tiny
%\caption{\autoref{Tab:flux} - continued}
\subcaption{Table 1 - continued}
 \begin{tabular}{rcccccccccccccccccccl}
  \hline\noalign{\smallskip}
       HIP   & B & V & J & H & K & w1m & w2m &  w3m &  w4m & f12 & f25& f60& f100 & $\chi^2$ \\
               & mag & mag & mag & mag & mag & mag & mag & mag & mag & Jy & Jy & Jy & Jy &  \\
       \hline\noalign{\smallskip}
     70090  &  4.02  &  4.05 &   4.11 &   3.98  &  4.07 &  4.063 &  3.851 &  4.092 &  3.914 &   0.953  &   0.313  &   0.171  &   1.233 & 3.46e-02 \\
    70344  &  7.80  &  7.21 &   6.12 &   5.87  &  5.79 &  5.816 &  5.714 &  5.773 &  5.274 &   0.185  &   0.089  &   0.182  &   0.586 & 1.89e-01 \\
    70952  &  6.53  &  6.10 &   5.26 &   5.09  &  5.01 &  5.447 &  5.194 &  5.965 &  5.151 &   0.349  &   0.084  &   0.344  &   0.525 & 1.78e+00 \\
    71075  &  3.23  &  3.04 &   2.65 &   2.57  &  2.51 &  1.378 &  1.247 &  2.621 &  2.470 &   3.937  &   0.980  &   0.366  &   0.496 & 1.87e+00 \\
    71284  &  4.83  &  4.47 &   3.56 &   3.46  &  3.34 &  3.478 &  3.146 &  3.507 &  3.444 &   1.606  &   0.413  &   0.140  &   0.452 & 1.78e+00 \\
    73049  &  5.37  &  5.32 &   5.22 &   5.22  &  5.13 &  5.173 &  5.010 &  5.193 &  5.031 &   0.332  &   0.155  &   0.190  &   1.342 & 5.66e-02 \\
      73145  &  8.07  &  7.88 &   7.60 &   7.56  &  7.52 &  7.514 &  7.523 &  6.922 &  4.273 &   0.156  &   0.186  &   0.684  &   1.324 & 6.53e-29 \\
    73473  &  4.91  &  4.91 &   4.71 &   4.71  &  4.62 &  4.602 &  4.405 &  4.681 &  4.584 &   0.623  &   0.282  &   0.141  &   0.974 & 5.57e-02 \\
    73512  & 10.17  &  9.13 &   7.19 &   6.69  &  6.58 &  6.577 &  6.555 &  6.572 &  6.527 &   0.123  &   0.122  &   0.130  &   0.405 & 9.28e-01 \\
    74596  &  5.34  &  5.28 &   5.09 &   5.15  &  5.11 &  5.094 &  4.926 &  5.150 &  4.939 &   0.348  &   0.100  &   0.142  &   0.601 & 4.97e-02 \\
    74946  &  2.88  &  2.87 &   2.52 &   2.53  &  2.53 &  1.597 &  1.539 &  2.731 &  2.732 &   3.372  &   0.763  &   0.416  &   5.406 & 1.78e+00 \\
    76127  &  4.01  &  4.14 &   4.43 &   4.47  &  4.43 &  4.412 &  4.133 &  4.394 &  4.228 &   0.623  &   0.182  &   0.162  &   0.452 & 3.43e-30 \\
    76267  &  2.25  &  2.22 &   2.25 &   2.39  &  2.21 &  0.942 &  0.935 &  2.263 &  1.940 &   5.807  &   1.686  &   0.705  &   0.869 & 3.12e-01 \\
    76375  &  8.60  &  7.65 &   6.01 &   5.62  &  5.50 &  5.698 &  5.325 &  5.511 &  5.466 &   0.211  &   0.070  &   0.142  &   0.614 & 1.11e+00 \\
    76635  &  8.05  &  7.50 &   6.50 &   6.25  &  6.18 &  6.202 &  6.130 &  6.201 &  5.985 &   0.199  &   0.114  &   0.131  &   0.936 & 7.64e-03 \\
    76736  &  6.53  &  6.45 &   6.30 &   6.34  &  6.27 &  6.270 &  6.219 &  6.160 &  5.010 &   0.263  &   0.250  &   0.482  &   4.657 & 4.30e+00 \\
    76829  &  5.05  &  4.64 &   4.02 &   3.73  &  3.80 &  3.681 &  3.087 &  3.650 &  3.521 &   1.425  &   0.693  &   0.661  &   2.368 & 4.87e+00 \\
    77163  &  5.61  &  5.57 &   5.47 &   5.46  &  5.43 &  5.402 &  5.258 &  5.428 &  4.997 &   0.330  &   0.124  &   0.283  &   0.920 & 8.59e+00 \\
    77542  &  7.21  &  7.11 &   6.87 &   6.86  &  6.82 &  6.061 &  6.095 &  4.701 &  1.861 &   0.533  &   1.819  &   5.335  &   3.548 & 1.78e+00 \\
    78554  &  4.89  &  4.82 &   5.01 &   4.66  &  4.62 &  4.647 &  4.455 &  4.653 &  4.574 &   0.577  &   0.145  &   0.190  &   0.749 & 3.26e+00 \\
    81126  &  4.19  &  4.20 &   4.42 &   4.34  &  4.05 &  4.173 &  3.903 &  4.175 &  4.099 &   0.937  &   0.270  &   0.265  &   0.651 & 4.49e+00 \\
    81641  &  5.77  &  5.77 &   5.75 &   5.79  &  5.74 &  5.824 &  5.712 &  5.792 &  5.340 &   0.207  &   0.111  &   0.226  &   1.189 & 3.55e-02 \\
    81800  &  7.02  &  6.48 &   5.44 &   5.17  &  5.15 &  5.168 &  4.994 &  5.166 &  5.023 &   0.321  &   0.073  &   0.116  &   0.640 & 1.02e-01 \\
    82405  &  6.72  &  6.32 &   5.46 &   5.29  &  5.18 &  5.188 &  5.019 &  5.202 &  5.123 &   0.381  &   0.262  &   0.295  &   1.148 & 8.22e-02 \\
    83480  &  6.89  &  6.70 &   6.30 &   6.25  &  6.22 &  6.190 &  6.092 &  6.190 &  6.048 &   0.156  &   0.104  &   0.321  &   2.245 & 1.96e-01 \\
    85157  &  5.93  &  5.70 &   5.24 &   5.21  &  5.18 &  5.178 &  5.028 &  5.167 &  4.018 &   0.356  &   0.212  &   0.536  &   1.086 & 2.54e-01 \\
    85537  &  5.65  &  5.41 &   4.81 &   4.88  &  4.80 &  4.777 &  4.572 &  4.802 &  4.599 &   0.572  &   0.258  &   0.366  &   2.093 & 2.38e-01 \\
    87108  &  3.79  &  3.75 &   3.59 &   3.66  &  3.62 &  3.677 &  3.356 &  3.646 &  3.118 &   1.417  &   0.525  &   1.269  &   3.733 & 1.21e-05 \\
    87558  &  6.20  &  5.77 &   5.21 &   4.83  &  4.71 &  4.737 &  4.453 &  4.717 &  4.576 &   0.544  &   0.250  &   0.380  &   1.265 & 3.00e+00 \\
    88399  &  7.47  &  7.01 &   6.16 &   6.02  &  5.91 &  5.758 &  5.703 &  5.799 &  4.940 &   0.208  &   0.158  &   0.647  &   2.631 & 2.83e+00 \\
    90185  &  1.76  &  1.79 &   1.73 &   1.77  &  1.77 & -0.704 & -0.165 &  1.763 &  1.681 &   8.460  &   2.047  &   0.586  &   2.616 & 7.25e-01 \\
    90936  &  6.65  &  6.22 &   5.42 &   5.28  &  5.20 &  5.206 &  5.053 &  5.224 &  5.082 &   0.292  &   0.139  &   0.504  &   1.029 & 1.32e-01 \\
    91262  &  0.03  &  0.03 &  -0.18 &  -0.03  &  0.13 & -2.034 & -2.082 &  0.017 & -0.157 &  41.530  &  11.310  &   9.568  &   7.912 & 9.42e-02 \\
    92024  &  4.98  &  4.78 &   4.38 &   4.25  &  4.30 &  4.335 &  4.065 &  3.624 &  2.355 &   1.518  &   1.092  &   0.306  &   0.991 & 8.61e-30 \\
    93542  &  4.71  &  4.74 &   4.72 &   4.96  &  4.75 &  4.794 &  4.526 &  4.723 &  3.792 &   0.577  &   0.306  &   0.321  &   1.119 & 6.98e-01 \\
    95261  &  5.05  &  5.03 &   5.10 &   5.15  &  5.01 &  5.059 &  4.822 &  4.725 &  3.261 &   0.481  &   0.491  &   0.495  &   0.435 & 1.45e-29 \\
    95270  &  7.52  &  7.04 &   6.20 &   5.98  &  5.91 &  5.887 &  5.810 &  5.894 &  3.952 &   0.164  &   0.248  &   1.855  &   1.719 & 1.51e-01 \\
    95619  &  5.65  &  5.66 &   5.67 &   5.66  &  5.68 &  5.677 &  5.559 &  5.612 &  4.579 &   0.190  &   0.199  &   0.412  &   0.942 & 1.79e+00 \\
    96468  &  4.28  &  4.36 &   4.44 &   4.42  &  4.48 &  4.549 &  4.256 &  4.593 &  4.511 &   0.609  &   0.166  &   0.265  &   1.804 & 8.80e-01 \\
    99273  &  7.66  &  7.18 &   6.32 &   6.09  &  6.08 &  6.062 &  5.882 &  5.995 &  4.056 &   0.111  &   0.339  &   0.711  &   0.328 & 1.96e-03 \\
    99473  &  3.17  &  3.24 &   3.29 &   3.28  &  3.29 &  3.390 &  3.024 &  3.435 &  3.350 &   1.745  &   0.447  &   0.204  &   0.941 & 1.78e+00 \\
   101612  &  5.04  &  4.75 &   4.28 &   4.02  &  4.04 &  4.060 &  3.582 &  4.045 &  3.865 &   1.016  &   0.264  &   0.522  &   1.798 & 3.38e-01 \\
   101769  &  4.07  &  3.64 &   2.87 &   2.76  &  2.64 &  1.584 &  0.844 &  2.634 &  2.518 &   4.069  &   0.921  &   0.229  &   1.066 & 8.24e-02 \\
    \noalign{\smallskip}\hline
\end{tabular}
\end{center}
\end{table}

\clearpage
\begin{table}
\begin{center}
%\tiny
%\caption{\autoref{Tab:flux} - continued}
\subcaption{Table 1 - continued}
 \begin{tabular}{rcccccccccccccccccccl}
  \hline\noalign{\smallskip}
       HIP   & B & V & J & H & K & w1m & w2m &  w3m &  w4m & f12 & f25& f60& f100 & $\chi^2$ \\
               & mag & mag & mag & mag & mag & mag & mag & mag & mag & Jy & Jy & Jy & Jy &  \\
       \hline\noalign{\smallskip}
   101800  &  5.47  &  5.42 &   5.41 &   5.37  &  5.30 &  5.324 &  5.134 &  5.326 &  4.968 &   0.251  &   0.124  &   0.177  &   1.077 & 8.79e-01 \\
   102409  & 10.28  &  8.81 &   5.44 &   4.83  &  4.53 &  4.499 &  3.999 &  4.312 &  4.137 &   0.760  &   0.300  &   0.270  &   0.687 & 1.13e-02 \\
   103752  &  6.46  &  6.36 &   6.13 &   6.11  &  6.05 &  6.056 &  5.921 &  6.085 &  6.121 &   0.146  &   0.103  &   0.244  &   0.638 & 2.07e-03 \\
   105570  &  5.22  &  5.16 &   5.09 &   4.98  &  4.92 &  4.885 &  4.699 &  4.966 &  4.793 &   0.459  &   0.251  &   0.274  &   1.925 & 3.40e-01 \\
   106741  &  7.59  &  7.19 &   6.38 &   6.25  &  6.18 &  6.192 &  6.066 &  6.185 &  5.927 &   0.159  &   0.113  &   0.199  &   0.520 & 7.66e-02 \\
   107022  &  7.83  &  7.07 &   5.73 &   5.41  &  5.27 &  5.309 &  5.121 &  5.306 &  5.209 &   0.315  &   0.107  &   0.174  &   0.987 & 1.23e-02 \\
   107412  &  7.13  &  6.69 &   5.87 &   5.69  &  5.59 &  5.573 &  5.452 &  5.629 &  5.481 &   0.249  &   0.194  &   0.228  &   0.491 & 1.02e-02 \\
   107649  &  6.17  &  5.57 &   4.72 &   4.31  &  4.24 &  4.116 &  3.817 &  4.156 &  4.065 &   0.805  &   0.216  &   0.245  &   1.000 & 5.01e+00 \\
   108809  &  7.13  &  6.63 &   5.70 &   5.44  &  5.39 &  5.338 &  5.180 &  5.393 &  5.251 &   0.287  &   0.152  &   0.137  &   0.680 & 1.33e-01 \\
    \noalign{\smallskip}\hline
\end{tabular}
\end{center}
\tablecomments{1.13\textwidth}{Col.(1): \hip\ identification.
%\tablecomments{1.25textwidth}{Col.(1): \hip\ identification.
Col.(2): B magnitude. Col.(3): V magnitude. Col.(4): J magnitude.
Col.(5): H magnitude.Col.(6): K magnitude.  Col.(7): \wise\ W1 magnitude. Col.(8): \wise\ W2 magnitude. Col.(9): \wise\ W3 magnitude. Col.(10): \wise\ W4 magnitude.Col.(11): \iras\ 12 $\mu$m flux density. Col.(12): \iras\ 25 $\mu$m flux density. Col.(13): \iras\ 60 $\mu$m flux density.
Col.(14): \iras\ 100 $\mu$m flux density. Col.(15): The minimum $\chi^2$ of SED fittings.
 }
\end{table}
%---------------------------------------------------------------------------------------

\clearpage
\begin{table}
\begin{center}
\caption[]{ The list of \wise\ 22 $\mu$m Excess Stars. }\label{Tab:WISE}
 \begin{tabular}{rcl}
  \hline\noalign{\smallskip}
       HIP   & Mid-IR excess   &  references \\
       \hline\noalign{\smallskip}
        5626  &   no  &        \\
      7345  &  yes  &   1,2,3,4  \\
     11847  &  yes  &   2,3,5 \\
     16449  &  yes  &    2    \\
     18437  &  yes  &    2    \\
     22226  &  yes  &    2,5  \\
     23451  &   no  &    3,5  \\
     24528  &  yes  &    2    \\
     26453  &  yes  &    2,5  \\
     26966  &  yes  &    2    \\
     27288  &   no  &         \\
     27321  &   no  &   1,3,4 \\
     28230  &  yes  &    2   \\
     36948  &   no  &    5  \\
     41307  &  yes  &        \\
     48541  &  yes  &    2   \\
     53524  &   no  &   3    \\
     53911  &  yes  &        \\
     55505  &  yes  &   1,3  \\
     61498  &   no  &   1,3  \\
     61782  &   no  &   1,4  \\
     73145  &   no  &   3,5  \\
     76736  &   no  &   3,5  \\
     85157  &   no  &        \\
     88399  &   no  &   3    \\
     92024  &  yes  &   4    \\
     93542  &  yes  &   2    \\
     95261  &  yes  &   1,2  \\
     95270  &  yes  &   2,3,5\\
     99273  &   no  &   3    \\
    116431  &   no  &   2    \\
    \noalign{\smallskip}\hline
\end{tabular}
\end{center}
\tablecomments{0.5\textwidth}{
Col.(1): \hip\ identification.
Col.(2): Check the 22 $\mu$m excess by \iras\ 100 $\mu$m background level lower than 5 MJy/sr or not, yes means the excess should be true and no means the excess may not be true.
Col.(3): References - 1. \citealt{fuj13}; 2. \citealt{wu13}; 3. \citealt{liu14}; 4. \citealt{ish17}; 5. \citealt{bal18}. }
\end{table}

\clearpage
\begin{table}
\begin{center}
\caption[]{ The Dust Properties of Group I Stars.} \label{Tab:groupi}
 \begin{tabular}{rcccc}
  \hline\noalign{\smallskip}
     HIP  &   T-Rhee(K)  &   f-Rhee      &   T(K)  &    f    \\
     \hline\noalign{\smallskip}
     4267   &    85   & 8.80e-05  &     73  &  7.13e-05 \\
     6686   &    85   & 5.90e-06  &    105  &  6.17e-06 \\
     8122   &    85   & 4.70e-04  &     66  &  3.40e-04 \\
     9570   &    85   & 1.00e-04  &     69  &  8.16e-05 \\
    11486   &    85   & 1.10e-04  &     85  &  1.13e-04 \\
    11847   &    85   & 1.70e-03  &     83  &  1.43e-03 \\
    13005   &    85   & 1.10e-03  &     52  &  7.56e-04 \\
    18437   &    85   & 2.60e-04  &    105  &  2.93e-04 \\
    18859   &    85   & 1.30e-04  &     94  &  1.38e-04 \\
    18975   &    85   & 8.90e-05  &     69  &  7.45e-05 \\
    20635   &    85   & 4.70e-05  &     81  &  4.33e-05 \\
    23451   &    85   & 5.40e-03  &     87  &  4.68e-03 \\
    25790   &    85   & 2.50e-04  &     55  &  1.21e-04 \\
    26966   &    85   & 2.00e-04  &     99  &  1.76e-04 \\
    34276   &    85   & 2.00e-04  &     73  &  1.72e-04 \\
    36906   &    85   & 4.30e-04  &     50  &  2.95e-04 \\
    39757   &    85   & 5.40e-06  &    128  &  1.03e-05 \\
    40938   &    85   & 3.50e-04  &     53  &  2.29e-04 \\
    42028   &    85   & 7.10e-05  &     59  &  5.16e-05 \\
    43970   &    85   & 1.00e-04  &     73  &  6.48e-05 \\
    44001   &    85   & 2.20e-04  &     66  &  1.25e-04 \\
    45758   &    85   & 2.70e-04  &     51  &  1.82e-04 \\
    48164   &    85   & 5.50e-04  &     63  &  2.97e-04 \\
    48541   &    85   & 4.80e-04  &     76  &  2.85e-04 \\
    51438   &    85   & 2.40e-05  &     79  &  2.49e-05 \\
    53524   &    85   & 1.50e-03  &     69  &  7.55e-04 \\
    56253   &    85   & 1.00e-04  &     61  &  6.66e-05 \\
    61960   &    85   & 6.20e-05  &     94  &  7.11e-05 \\
    64375   &    85   & 3.90e-04  &     64  &  3.04e-04 \\
    69682   &    85   & 2.10e-03  &     62  &  1.43e-03 \\
    70344   &    85   & 3.80e-04  &     80  &  3.36e-04 \\
    73049   &    85   & 7.40e-05  &     67  &  5.67e-05 \\
    73512   &    85   & 1.20e-03  &     54  &  7.25e-04 \\
    76635   &    85   & 3.90e-04  &     64  &  2.96e-04 \\
    82405   &    85   & 3.00e-04  &     53  &  1.81e-04 \\
    83480   &    85   & 4.30e-04  &     52  &  2.14e-04 \\
    87108   &    85   & 7.80e-05  &     81  &  7.68e-05 \\
    87558   &    85   & 2.50e-04  &     69  &  2.00e-04 \\
    95619   &    85   & 2.00e-04  &     88  &  1.71e-04 \\
    99473   &    85   & 6.60e-06  &     92  &  7.22e-06 \\
   103752   &    85   & 2.50e-04  &     52  &  1.56e-04 \\
   105570   &    85   & 8.80e-05  &     68  &  6.86e-05 \\
   106741   &    85   & 4.00e-04  &     61  &  2.01e-04 \\
   110867   &    85   & 7.10e-04  &     59  &  5.13e-04 \\
   116431   &    85   & 7.40e-04  &     84  &  5.23e-04 \\
    \noalign{\smallskip}\hline
\end{tabular}
\end{center}
\tablecomments{0.6\textwidth}{Col.(1): \hip\ identification.
Col.(2): Dust temperatures of Rhee.
Col.(3): Fractional luminosities of Rhee.
Col.(4): The revised dust temperatures of this paper.
Col.(5): The revised fractional luminosities of this paper. }
\end{table}

\clearpage
\begin{table}
\begin{center}
\caption{ The Dust Properties of Group II Stars.} \label{Tab:groupii}
 \begin{tabular}{rcccc}
  \hline\noalign{\smallskip}
     HIP &   T-Rhee(K)  &   f-Rhee      &   T(K)  &    f    \\
     \hline\noalign{\smallskip}
      746  &    120 & 2.50e-05  &    109 &  1.85e-05 \\
     1185  &     40 & 4.30e-04  &     36 &  2.67e-04 \\
     5626  &     75 & 1.50e-04  &     83 &  1.40e-04 \\
     6878  &     45 & 2.10e-04  &     50 &  1.88e-04 \\
     7345  &     80 & 7.90e-04  &     76 &  7.90e-04 \\
     7805  &     70 & 3.70e-04  &     74 &  2.69e-04 \\
     7978  &     65 & 4.20e-04  &     68 &  3.63e-04 \\
     8241  &     75 & 6.40e-05  &     76 &  7.03e-05 \\
    10054  &     60 & 8.70e-05  &     66 &  5.20e-05 \\
    10670  &     75 & 7.20e-05  &     74 &  5.57e-05 \\
    11360  &     65 & 5.10e-04  &     68 &  4.74e-04 \\
    12361  &     40 & 5.90e-04  &     37 &  3.22e-04 \\
    12964  &     55 & 2.00e-04  &     54 &  1.27e-04 \\
    13141  &     55 & 6.40e-05  &     83 &  4.81e-05 \\
    14576  &    250 & 1.70e-05  &    138 &  7.79e-06 \\
    15197  &     95 & 2.50e-05  &     79 &  4.61e-05 \\
    16449  &     60 & 4.90e-04  &     66 &  4.71e-04 \\
    16537  &     40 & 8.30e-05  &     53 &  7.52e-05 \\
    19893  &     80 & 2.30e-05  &     74 &  3.81e-05 \\
    21604  &     75 & 3.80e-04  &     54 &  1.89e-04 \\
    22226  &     65 & 8.80e-04  &     70 &  6.51e-04 \\
    22439  &     40 & 2.30e-04  &     62 &  1.21e-04 \\
    22845  &     80 & 8.40e-05  &     88 &  8.24e-05 \\
    24528  &    100 & 1.70e-04  &     94 &  1.82e-04 \\
    25197  &    120 & 7.00e-05  &     86 &  5.06e-05 \\
    26453  &     90 & 2.80e-04  &     97 &  3.21e-04 \\
    27072  &     90 & 7.70e-06  &    133 &  1.93e-05 \\
    27288  &    220 & 1.30e-04  &    175 &  8.09e-05 \\
    27321  &    110 & 2.60e-03  &    109 &  2.65e-03 \\
    27980  &     70 & 2.80e-03  &     38 &  2.62e-03 \\
    28103  &    185 & 2.00e-05  &    114 &  1.29e-05 \\
    28230  &     90 & 6.10e-04  &     82 &  4.35e-04 \\
    32480  &     60 & 8.90e-05  &     82 &  1.29e-04 \\
    32775  &     45 & 1.60e-04  &     54 &  1.07e-04 \\
    33690  &     80 & 2.00e-04  &     82 &  2.31e-04 \\
    34819  &     45 & 1.00e-04  &     72 &  7.69e-05 \\
    35550  &     60 & 8.90e-06  &     82 &  1.13e-05 \\
    36948  &     60 & 2.60e-03  &     63 &  2.31e-03 \\
    41152  &     80 & 5.20e-05  &     85 &  5.23e-05 \\
    41307  &    130 & 4.10e-05  &    128 &  4.36e-05 \\
    42430  &     80 & 3.20e-05  &    119 &  5.21e-05 \\
    51658  &     40 & 1.10e-04  &     37 &  6.79e-05 \\
    52462  &     45 & 6.70e-04  &     50 &  5.43e-04 \\
    53910  &    120 & 1.20e-05  &     94 &  1.22e-05 \\
    55505  &    160 & 1.10e-01  &    136 &  6.59e-02 \\
    56675  &     50 & 1.40e-04  &     65 &  6.84e-05 \\
    57632  &    160 & 4.20e-05  &     88 &  1.13e-05 \\
        60074  &     55 & 9.50e-04  &     57 &  7.47e-04 \\
      \noalign{\smallskip}\hline
\end{tabular}
\end{center}
\end{table}

%\clearpage
\begin{table}
\begin{center}
%\subcaption{\autoref{Tab:groupii} - continued}
\subcaption{Table 4 - continued}
 \begin{tabular}{rcccc}
  \hline\noalign{\smallskip}
     HIP &   T-Rhee(K)  &   f-Rhee      &   T(K)  &    f    \\
     \hline\noalign{\smallskip}
    61174  &    180 & 1.20e-04  &    156 &  1.05e-04 \\
    61498  &    110 & 4.40e-03  &    108 &  4.22e-03 \\
    61782  &    130 & 2.50e-03  &    119 &  2.41e-03 \\
    63584  &    100 & 1.00e-04  &     55 &  4.43e-05 \\
    64921  &     80 & 3.40e-04  &     55 &  2.10e-04 \\
    68101  &     45 & 2.50e-04  &     71 &  1.93e-04 \\
    68593  &     60 & 1.40e-04  &     62 &  2.19e-04 \\
    69732  &    100 & 5.20e-05  &     84 &  4.59e-05 \\
    70090  &    120 & 2.10e-05  &     89 &  1.53e-05 \\
    71075  &     55 & 1.00e-05  &     69 &  8.35e-06 \\
    73145  &     90 & 3.10e-03  &     91 &  2.95e-03 \\
    73473  &    150 & 7.20e-05  &     69 &  1.79e-05 \\
    74596  &     65 & 3.30e-05  &     74 &  2.97e-05 \\
    76127  &     75 & 2.00e-05  &     99 &  7.85e-06 \\
    76267  &    190 & 2.40e-05  &    127 &  1.97e-05 \\
    76375  &     29 & 7.90e-04  &     28 &  3.97e-04 \\
    76736  &    140 & 1.20e-04  &     79 &  3.99e-04 \\
    76829  &     75 & 1.20e-04  &     77 &  1.29e-04 \\
    77163  &     40 & 1.40e-04  &     72 &  1.16e-04 \\
    78554  &     45 & 4.60e-05  &     76 &  3.73e-05 \\
    81126  &     80 & 3.00e-05  &     80 &  2.57e-05 \\
    81641  &     95 & 1.20e-04  &     74 &  9.24e-05 \\
    81800  &     55 & 8.30e-05  &     66 &  9.49e-05 \\
    85157  &     90 & 2.70e-04  &     92 &  2.14e-04 \\
    85537  &     70 & 6.80e-05  &     63 &  6.52e-05 \\
    88399  &     70 & 1.00e-03  &     72 &  6.40e-04 \\
    90185  &    100 & 4.50e-06  &    115 &  4.80e-06 \\
    90936  &     50 & 4.60e-04  &     52 &  2.25e-04 \\
    91262  &     80 & 2.10e-05  &     76 &  1.77e-05 \\
    92024  &    320 & 8.10e-04  &    269 &  6.18e-04 \\
    93542  &    120 & 9.70e-05  &    116 &  8.99e-05 \\
    95261  &    150 & 2.10e-04  &    126 &  2.17e-04 \\
    95270  &     75 & 3.50e-03  &     75 &  3.38e-03 \\
    96468  &     60 & 3.50e-05  &     70 &  2.26e-05 \\
    99273  &     95 & 1.40e-03  &     91 &  1.79e-03 \\
   101612  &     65 & 1.10e-04  &     71 &  1.00e-04 \\
   101769  &    130 & 1.60e-05  &    116 &  1.06e-05 \\
   101800  &    100 & 3.90e-05  &     83 &  6.34e-05 \\
   102409  &     50 & 3.60e-04  &     60 &  3.99e-04 \\
   107022  &     80 & 2.90e-04  &     55 &  2.35e-04 \\
   107412  &     55 & 2.70e-04  &     58 &  2.33e-04 \\
   107649  &     55 & 1.20e-04  &     84 &  1.12e-04 \\
   108809  &     75 & 7.30e-05  &     67 &  1.33e-04 \\
   109857  &     65 & 1.60e-04  &     67 &  1.39e-04 \\
   111278  &     55 & 9.30e-05  &     47 &  8.48e-05 \\
   113368  &     65 & 8.00e-05  &     57 &  5.72e-05 \\
   114189  &     50 & 2.30e-04  &     69 &  1.70e-04 \\
    \noalign{\smallskip}\hline
\end{tabular}
\end{center}
\tablecomments{0.5\textwidth}{The explanation of each column is same to Table 3. }
\end{table}

\label{lastpage}

\end{document}